\documentclass[aps,prl,reprint,superscriptaddress]{revtex4-1}
\usepackage{graphicx}
\usepackage{epstopdf}
\usepackage{amsmath}
\usepackage{footnote}
\usepackage[latin1]{inputenc}

\begin{document}

\title{Interference  between magnetic field and cavity modes in an extended Josephson junction}

\author{V. Humbert}
\affiliation{Laboratoire de Physique des Solides, UMR8502-CNRS, University Paris-Sud, 91405 Orsay Cedex, France.}
\author{M. Aprili}
\email[]{aprili@lps.u-psud.fr}
\affiliation{Laboratoire de Physique des Solides, UMR8502-CNRS, University Paris-Sud, 91405 Orsay Cedex, France.}
\author{J. Hammer}
\altaffiliation[Current address: ]{Max-Planck-Institut f\"{u}r Quantenoptik, D-85748 Garching, Germany.}
\affiliation{Institute for Experimental and Applied Physics, University of Regensburg, 93040 Regensburg, Germany.}

\date{\today}

\begin{abstract}

An extended Josephson junction consists of two superconducting electrodes that are separated by an insulator and it is therefore also a microwave cavity. The superconducting phase difference across the junction determines the supercurrent as well as its spatial distribution. Both, an external magnetic field and a resonant  cavity intrafield produce a spatial modification of the superconducting phase along the junction. The interplay between these two effects leads to interference in the critical current of the junction and allows us to continuously tune the coupling strength between the first cavity mode and the Josephson phase from $1$ to $-0.5$\,. This enables static and dynamic control over the junction in the ultra-strong coupling regime.

 \end{abstract}
%
%

\maketitle

 A Josephson junction can be 
described as a two level system, at sufficiently low temperature, due to the non-linearity of the Josephson coupling. The strong coupling of a Josephson junction to an on-chip microwave superconducting resonator with small losses (i.e. quality factor higher than $10^4$) has led to the emergence of the new field of circuit quantum electrodynamics (CQED)  \cite{wallraff}. We note that an  extended Josephson junction in which the two superconductors are coupled through an insulating barrier, is at the same time a non-linear Josephson oscillator $and$ a microwave cavity \cite{likharev}. Neglecting the Josephson effect, the eigenfrequencies of the electromagnetic modes are given by  $\nu_n = k_n \cdot c_s/2\pi$, with $k_n = n \cdot \pi / L$ where L is the junction length (see Fig. 1(b)) and  $c_s$ the Swihart velocity \cite{swihart}. As the Josephson current and the microwave field are both localized in the insulator, extended junctions intrinsically form a microwave cavity enclosing a material resonance, the superconducting oscillator. 
Therefore, not only, the Josephson plasma frequency and the mode frequencies can be made, at low temperature, much larger than damping as required for strong coupling, but more importantly, the vacuum Rabi frequency, i.e. the photon exchange rate between the microwave cavity and the Josephson oscillator, can be as large as a fraction of the first cavity mode eigenfrequency, so that in fact the system is in the ultra-strong coupling limit \cite{ciuti}.
This can be seen from the Hamiltonian of the junction \cite{fistul}, $H=H_J+H_c+H_{int}$ where $H_J$ and $H_c$ are the Josephson and the cavity Hamiltonian, respectively, while $H_{int}= - (h \nu_p)^2 / h \nu_n g_n N_c \varphi_0$ describes the interaction between the cavity modes and the Josephson phase. Here $\nu_p$ is the Josephson plasma frequency, $N_c$ the cavity photon number and $\varphi_0$ the macroscopic phase difference across the junction \cite{note}. The interaction thus provides an intrinsically non-linear coupling which is formally equivalent to radiation-pressure interaction in optomechanics \cite{aspelmeyer}, \cite{marquardt}. The coupling constant $g_n$ is given by  

\begin{equation}
\label{S0}
\begin{split}
g_n= 2\int_{-\frac{L}{2}}^{\frac{L}{2}}\frac{dx}{L}& \sin[k_n x]^2 \cos [k_H x]=\frac{\sin[\frac{\pi \phi}{\phi_0}]}{\frac{\pi \phi}{\phi_0}}\cdot h_n(\phi)\\ \textmd {with~ } &h_n(\phi)=1+\frac{(\phi/\phi_0)^2\cos(n\,\pi)}{n^2-(\phi/\phi_0)^2}
\end{split}
%
\end{equation}

where $k_H=2\pi \phi/(\phi_0 L)$, with $\phi$ the magnetic flux in the junction and $\phi_0 = 2e/h$ the flux quantum \cite{fistul}. At zero applied magnetic field $g_1=1$, therefore extended Josephson  junctions  provide an appealing system to investigate ultra-strong coupling between the superconducting phase and photons \cite{gross} with small aging factors from the electromagnetic environment. This regime is difficult to achieve in optical cavities \cite{walther}, but recently has been obtained in a solid-state semiconductor system \cite{gunter}. Furthermore it is worth noting, that the coupling is statically and dynamically tunable. In fact, $g_1$ can be continuously changed from 1 to -0.5 by the external magnetic field. 

In this paper we report on the magnetic field dependence of the coupling strength between the Josephson phase and the first cavity mode. Here $g_1$ is obtained by measuring the Josephson critical current as a function of the applied magnetic field for a microwave radiation frequency either resonant or non-resonant with the first cavity mode.  Thus the magnetic field dependence of  $g_1$, corresponds physically to the interference in the Josephson critical current between the phase differences produced by the intracavity  and  the magnetic fields. 

The critical current of a planar rectangular Josephson junction shows a Fraunhofer pattern as a function of the applied magnetic field \cite{rowell} (see Fig. 1(c)). This originates from the phase difference, $\varphi_H$, created by the magnetic flux through the junction \cite{barone}. 
If self-screening of the applied magnetic field  (i. e.  $\lambda_L >> L$\,where $\lambda_L$ is the Josephson penetration depth) is neglected, the phase difference accumulated along the junction is  $\varphi_H$=$k_H$x. Here $d$ is magnetic penetration depth \cite{barone}. Therefore, the critical current, $I_c$ through the junction is given by \cite{barone} $I_c =I_{c0} \lvert \frac {\sin (\pi \phi/\phi_0)}{ \pi \phi/\phi_0}\rvert $. 

In presence of the n-th resonant mode due to microwave excitation, the phase difference produced by the electromagnetic field, $\varphi_{RF}$, has to be added to $\varphi_H$. The total phase difference is $\varphi$=$\varphi_H$+$\varphi_{RF}$+$\varphi_0$ where $\varphi_{RF} =a_n\cdot Re (e^{i 2\pi \nu _n t})  \sin(k_nx)$ is obtained by integrating the second Josephson equation with $a_n= 2 e V_{RF}/(\hbar \nu _n)$. The critical current through the junction after integration of the first Josephson equation over time and space becomes \cite{fistul} 

\begin{equation}
\label{S2}
I_c = I_{c0}\, \lvert \frac {\sin( \pi \phi/\phi_0)}{\pi \phi/\phi_0}\rvert \cdot\left(1 -\frac{a_n^2}{4}\cdot h_n(\phi)   \right)
\end{equation}
Therefore the intracavity field of each mode contributes differently to the diffraction pattern. The second term in eq. \ref{S2} gives the magnetic field dependence of the coupling strength to the n-th resonant mode, $g_n$, and it accounts for the interference between $\varphi_H$\,and $\varphi_{RF}$. For simplicity we consider only the first mode, resulting in a magnetic field dependend deviation of the critical current according to

\begin{equation}
\label{S3}
\Delta I_c=\left(\frac{2 e V_{RF}}{\hbar \nu _1}\right)^2\, \lvert \frac {\sin( \pi \phi/\phi_0)}{\pi \phi/\phi_0}\rvert\cdot \frac{1-2(\phi/\phi_0)^2}{4(1-(\phi/\phi_0)^2)}
\end{equation}

Let us make a few remarks about eq. \ref{S3}. First, since the phase-intrafield coupling is not linear, $\Delta I_c$ as a function of the magnetic flux is not equivalent to a normalization of the critical current and/or of magnetic quantum flux in the junction. Second $\Delta I_c$ changes sign at $0.7\,\phi_0$, meaning that the overall effect of microwave radiation is to decrease the critical current through the junction for $\phi < 0.7 \phi_0$, while it is to increase $I_c$ for $\phi > 0.7 \phi_0$. This is contrary to the common belief that microwave fields always reduce the Josephson critical current in the adiabatic approximation, i.e. when phase-photon coupled dynamics is not taken into account \cite{hammer}. Finally from the Josephson current-phase relation $I=I_c \sin(\varphi)$  we observe that there is a small flux range just above $\phi_0$ in which the macroscopic phase difference through the junction changes from $\pi$\,to $0$ under resonant microwave irradiation at frequency $\nu _1$. 
This is because close to $\phi_0$, the effect of the interference term $\Delta I_c$ is equivalent to a small shift in the Fraunhofer pattern. In long Superconductor/Normal/Superconductor Josephson  junctions microwave induced changes in the current-phase relation have been proposed \cite{yip} and observed \cite{jedema}, based on a completely different mechanism; namely the microwave pumping produces a strong out-of-equilibrium quasiparticle distribution in the Andreev bound states in the normal metal.

\begin{figure}[t]
		\includegraphics[width=8.6cm]{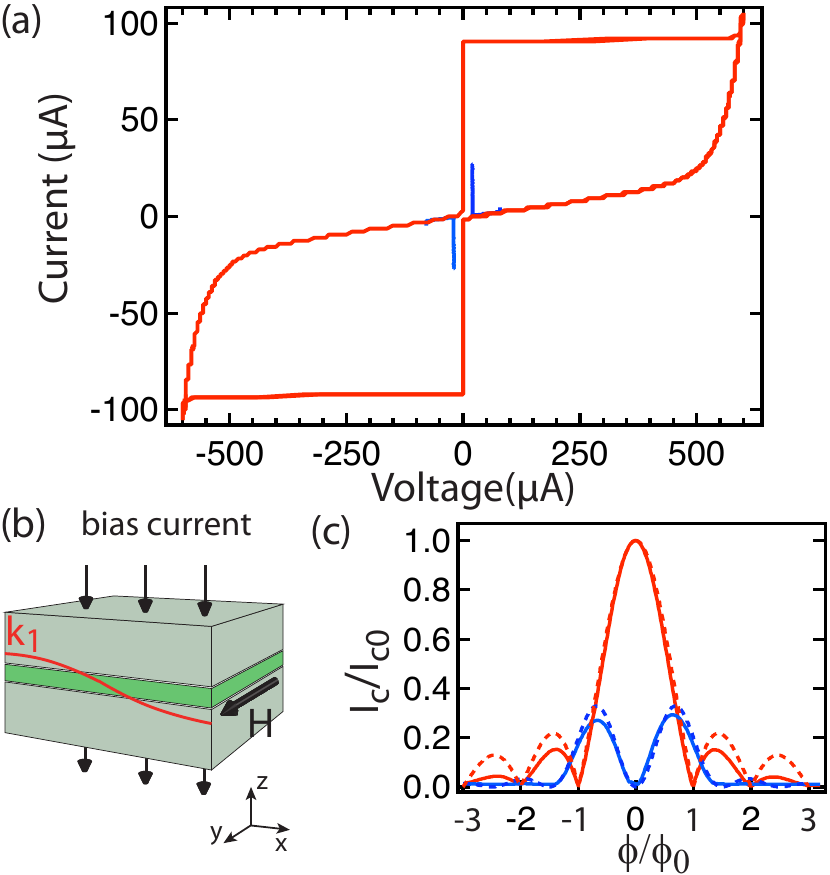}
			\caption{\label{fig:FIG2}(a) Hysteretic current-voltage characteristic (red line) of extended Josephson junction, taken at $600$\,mK. The first Fiske resonance at $\rm{V}=15µV$ (blue line) is taken at $\phi=0.7\, \phi_0$. (b) Sketch of extended Josephson junction. Electric field distribution of the first resonant mode $k_1$ is indicated in red. (c) Fraunhofer pattern of $I_c$ (data: red line, theory: red dashed line), and first Fiske resonance $I_{c1}$ (blue line) as a function of applied (in plane) magnetic field with theoretical curve (blue dashed line) given by eq. \ref{S4}.
}\end{figure}

\textit{Experiment.} We used  superconductor/insulator/\-ferromagnet/superconductor (SIFS) Josephson junctions consisting of $\rm Nb(150 nm)/Al_2O_3/PdNi(d_F)/\-Nb(50 nm)$ in a cross strip geometry (see Fig. 1(b)). The ferromagnetic thin layer reduces the Josephson coupling and the  phase relaxation time at the working temperature of 600mK. The weak ferromagnet PdNi contains $10 \%$ of Ni, has a Curie temperature of around 150 K, and its thickness $d_F$ varies between $50 \, \rm{\AA}$ and $100 \, \rm{\AA}$ \cite{kontos}. We also fabricated non-ferromagnetic (SIS) junctions without the PdNi layer, in order to verify that the thin ferromagnetic layer has no other effect than reducing the critical current. The fabrication details are given elsewhere \cite{kontos}. The critical current of ferromagnetic junctions is between 10 $\mu A$ and $130 \, \mu A$, the normal resistance $R_n \sim 0.2\, \Omega$, and the critical temperature around $8.2\,$K. The quasiparticle resistance of the junctions was measured to be 29 $\Omega$. The junction area is $0.7 \times 0.7 \, \rm mm^2$, the capacitance, $C$,  is  $30$\,nF \cite{hammer} making phase dynamics underdamped  \cite{barone}. The magnetic field $H$ is applied in the $y$-direction.  A $\mu$-metal shield ensures a negligible residual magnetic field in the one-shot $^3He$ cryostat.

\begin{figure}[t]
		\includegraphics[width=8.6cm]{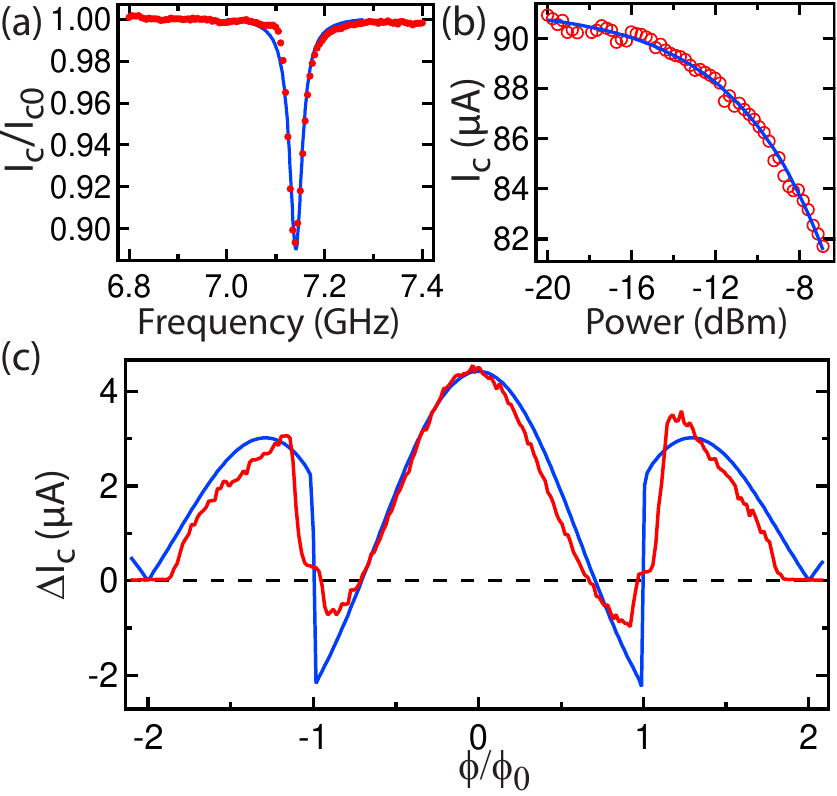}
			\caption{\label{fig:FIG2}(a) Josephson spectroscopy of first cavity mode: Resonance of normalized critical current as function of microwave frequency (red dots) at $\nu_1=7.18$\,GHz with quality factor $Q_c=250$ from Lorenzian fit (blue line). (b) Suppression of critical current $I_c$ as function of injected microwave power. Blue line corresponds to theoretical fit. (c) Deviation of the critical current $\Delta I_c$ for resonant ($\nu=7.185$\,GHz) and non-resonant ($\nu=6.950$\,GHz) microwave excitation. The blue line is obtained from theoretical predictions.
}\end{figure}

In Fig. 1(a) we show the current-voltage ($IV$) characteristic measured at zero applied magnetic field. The data follows a hysteretic $IV$ characteristic with the retrapping current practically zero, as expected for strongly underdamped Josephson junctions. The resonance at $V_1= 15 \mu$V (blue line) is the first Fiske step  \cite{fiske}. When a finite DC-voltage appears across the junction, the cavity modes are resonantly excited at $V_n^{DC}  = \frac{h}{2e} \nu _n $, and mix with the AC-Josephson current giving rise to finite DC-resonances \cite{fiske}. The first Fiske step shown in Fig. 1(a) has been recorded separately for an applied magnetic field  of $\phi=0.7\,\phi_0$ which maximizes the step amplitude. From  $V_1= 15 \mu$V we obtain a Swihart velocity $\tilde{c}=0.037 c$. In Fig. 1(a) we present only the first Fiske step but higher order steps (not shown) are also observed \cite{hammer}. We verify that the resonance at  $V_1= 15 \mu$V corresponds to the first Fiske resonance by measuring its magnetic field dependence, $I_{c1}(\phi)$, as reported in Fig. 1(c) (blue line). For the current amplitude of the first Fiske step one obtains from theory \cite{kulik}

\begin{equation}
\label{S4}
I_{c1}(\phi) =b\cdot I_{c0}\left( \frac{4 (\phi/\phi_0)}{2 (\phi/\phi_0) + 1} \frac{\sin(\pi(\phi/\phi_0-0.5))}{\pi(\phi/\phi_0-0.5)}  \right)^2
\end{equation}

as experimentally observed. Equation \ref{S4} is plotted in Fig.1(c) as a blue dashed line. Here the numerical constant $b=0.275$ is in the limit of high cavity quality factor $Q_c$. In Fig. 1(c) the critical current is also shown as a function of the applied magnetic field. $I_c$  follows the Fraunhofer pattern (red dashed line) as described above. Smaller secondary maxima indicate a larger current density in the center of the junction.

We now focus on the effect of the intracavity field on the Josephson switching current, i.e. on the maximum superconducting current  in the junction (at zero voltage bias), before it switches to the dissipative state, which corresponds to a finite voltage across the junction.
In our junctions at 600mK the switching current represents $I_c$ within one per cent \cite{petkovic}. The variation of the critical current as function of the microwave frequency for a fixed microwave injected power of $-15$\,dBm is shown in Fig. 2(a). We report the microwave power provided by the source and not the actual power arriving at the sample. The coupling constant is obtained below.  As microwaves are absorbed only at $\nu =\nu_n$, the critical current is suppressed only at resonance. This allows a fine spectroscopy of the cavity modes. We observed the first mode at $7.18$\,GHz as expected from the value measured for the first Fiske step ($15 \mu$V correspond to $7.5$\,GHz). From the Lorenzian fit (see Fig. 2(a)), we obtain the cavity quality factor $Q_c=250$. 

The quality factor is limited by dissipation. If dissipation is only due to quasiparticle tunneling, $Q_c$ would be given by $\omega_1 R_{qp} C$ (about $4\cdot10^4 $ in our junctions), where $R_{qp}$ is the tunneling quasiparticle resistance. Nevertheless, it has been shown \cite{qualityfactor} that  at  high frequency, surface losses in the electrodes and dielectric losses in the insulator are more important than quasiparticle tunneling and they both substantially reduce $Q_c$.

The microwave induced suppression of the critical current at resonance is given by $1-(\alpha a_1)^2$ \cite{shapiro}, where $\alpha$ is the coupling constant between the microwave line and the Josephson junction. In Fig. 2(b) we present $I_c$ versus microwave power at resonance, i.e. for $\nu = 7.18$\,GHz.
From the theoretical fit (blue line) we get $\alpha = 5\cdot10^{-5}$. This shows that the Josephson junction is very weakly coupled to the microwave circuit.
	
We then sweep the magnetic field and for each value of the applied field, we measure the difference in the critical current $\Delta I_c$\,, by substracting $I_c$ at two microwave frequencies $6.950$\,GHz and 7.185GHz for $-15$\,dBm, corresponding to $-235$\,MHz  and  $0$\,MHz detuning from the first cavity mode, respectively. Note that $235$\,MHz is much larger than the cavity bandwidth. $\Delta I_c$ as a function of the magnetic flux in the junction is reported in Fig. 2(c) (red line). Here $\Delta I_c$\,is the interference term between the intracavity and the magnetic field, described by eq. \ref{S3}, which is also shown in  Fig.2 (c) (blue line).  Equation \ref{S3} accounts well for the experimental data. In particular, we observe the change in sign of $\Delta I_c$  at $0.7\,\phi_0$ as predicted. The jump at about $\phi_0$ corresponds to the microwave induced $\pi$-$0$ transition in the macroscopic phase difference.

\begin{figure}[t]
		\includegraphics[width=8.6cm]{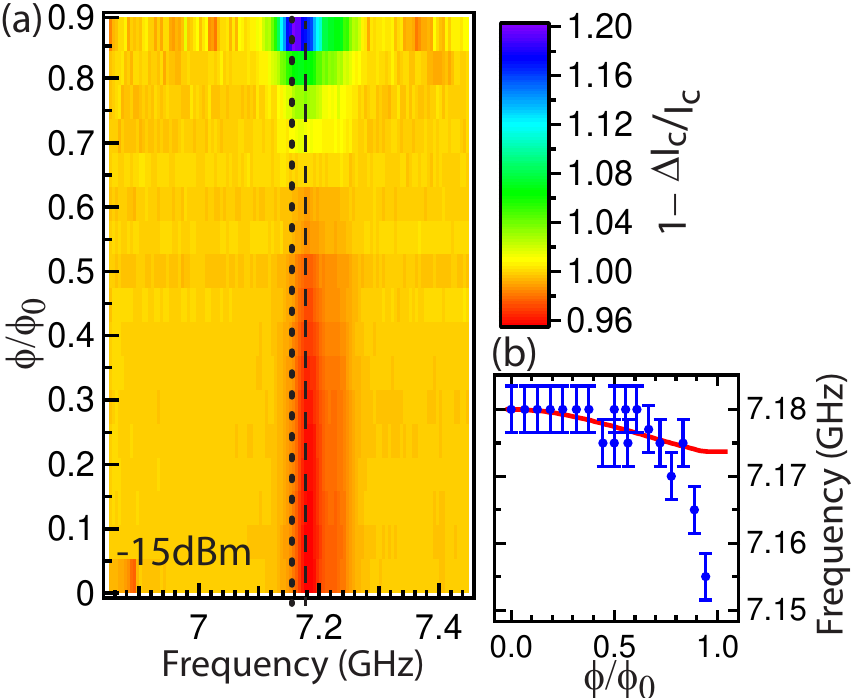}
			\caption{\label{fig:FIG2}(a) Colormap showing normalized deviation of critical current $1-\Delta I_c/I_c$\,as a function of microwave frequency and applied magnetic field. (b) Variation of the microwave resonant frequency for applied magnetic field, deduced from (a). The red line corresponds to the theoretically expected variation of $\nu_1$ as a function of an applied magnetic field.
}\end{figure}

The change in sign of $\Delta I_c$ can be seen more clearly in Fig. 3(a) where we plot the normalized critical current as a function of the microwave frequency and the applied magnetic field. The microwave power is $-15$\,dBm. Blue (red) corresponds to an increase (decrease) of the critical current. We observe that the frequency of the first resonance is slightly reduced by about $25$\,MHz, when $\Delta I_c$ changes sign, i.e. for $\phi > 0.7\,\phi_0$. The resonance frequency at $\phi =0$\,and $\phi=0.9\,\phi_0$ are marked by a dashed and a dotted line respectively in Fig.3(a). This magnetic field induced shift in the resonance frequency, $\Delta \nu_1$, is smaller than, but comparable to the cavity bandwidth and it explains the difference between data and theory in Fig. 2(c) for $0.7<(\phi/\phi_0) <1$. In fact, in this field range because of $\Delta \nu_1$, the value of  $\Delta I_c$\,measured as the difference in the critical current at two microwave frequencies $6.950$\,GHz and $7.185$\,GHz is underestimated, as can be seen in Fig.2(c). 
In  Fig.3(b) we present the value of $\Delta \nu_1$\,obtained from the data in Fig.3(a) as a function of the applied magnetic field.  We verified that $\Delta \nu_1$\,is independent on the microwave power. Due to the Josephson coupling the dispersion of the electromagnetic waves in the junction is  not linear, the cavity resonance frequencies become \cite{likharev}: $\nu_n = \sqrt{ \nu_p^2 +( k_n c_s)^2}$, where $\nu_p= 1/2\pi \sqrt{I_c /C\phi_0}$. The red line accounts for $\Delta\nu_1$\,when the magnetic field dependence of  the plasma frequency is taken into account. As clear from Fig. 3(b), the correction to $\nu_1$\,expected from the Josephson coupling is too small to explain the experimental changes observed in $\nu_1$. Therefore the shift in the resonance frequency $\nu_1$\,is likely related to the non-linear dependence of the kinetic inductance of the electrodes on the magnetic field induced screening. This is of course a small correction but measurable because of the high quality factor of the cavity. Thus it is interesting to point out that the measurement of $\nu_1$\,is a very effective way to directly determine the kinetic inductance at finite frequency (GHz regime) of complex superconducting based multilayers.

In conclusion we have observed that the phase difference produced by an applied magnetic field together with the phase difference caused by the cavity intrafield due to microwave radiation, interfere in extended Josephson junctions. This interference changes the  Fraunhofer pattern of the critical current. Moreover we have noticed that these types of junctions allow us to enter in the ultra-strong coupling limit between the Josephson oscillator and the cavity eigenmodes. The coupling strength is magnetic field dependent and it follows exactly the interference term measured with and without microwave radiation.

\begin{acknowledgments}
We thank  J. Gabelli, R. Gross, B. Reulet and A. Ustinov and  for valuable discussions. We are indebted to C.H.L. Quay for a critical reading of the manuscript.
\end{acknowledgments}


\begin{thebibliography}{1}
\bibitem{note} In our junctions the ratio $R=\nu_p /\nu_1$ is about 0.05, however, it is technically possible to achieve $R$ of the order of the unity.
\end{thebibliography}
 
\end{document}